\documentclass{iau} 
\usepackage{graphicx}

\title[Post-AGB nebular studies] 
{Post-AGB nebular studies}

\author[Eric Lagadec]   
{Eric Lagadec$^1$}
\affiliation{$^1$Universit\'e C\^ote d'Azur, Observatoire de la C\^ote d'Azur, CNRS, Lagrange, France \\ email: {\tt elagadec@oca.eu}}

\pubyear{2016}
\volume{323}  
\jname{Planetary Nebulae: Multi-Wavelength Probes of Stellar and Galactic Evolution}
\editors{X. Liu, L. Stanghellini, and A. Karakas, eds.}
\begin{document}

\maketitle

\begin{abstract}
This review presents the latest advances in the nebular studies of post-AGB objects. Post-AGB stars are great tools to  test nucleosynthesis and evolution models for stars of low and intermediate masses, and  the evolution of dust in harsh environment. I will present the newly discovered class of post-RGB stars, formed via binary interaction on the RGB.
Binary systems can also lead to the formation of  two class  of aspherical post-AGB, the  Proto-Planetary Nebulae and  the naked post-AGBs (dusty RV Taus , a.k.a. Van Winckel's stars). 
\end{abstract}

\firstsection 
\section{Introduction}
In this review, I present the most recent results based on the study
of post-AGB stars. For a thorough review of  the post-AGB objects, I
refer the reader to the very nice work by Hans van Winckel (\cite[Van
Winckel (2003)]{Vanwinckel2003}).

Post-AGB stars, as it can be guessed from their names, are evolved stars
of low to intermediate initial masses, in
the evolutionary phase after the Asymptotic Giant Branch.
The beginning of the post-AGB phase is usually defined as when the
envelope of the AGB star gets detached. The post-AGB phase ends when
this envelope gets ionised by the central star, which has shrunk
and became hotter. Observationally, post-AGB stars are often
identified by a double-peaked spectral energy distribution, with a
peak in the optical due to the contribution from stellar emission,
and a peak at longer wavelengths, in the infrared, due to cool dust in
the detached envelope. We will see in this review that this
observational identification is changing.

The advances in instrumentation, with new instruments/telescopes such as ALMA, SPHERE/VLT and the VLTI enable us to probe deep in the core of post-AGB objects and study them intensively. We can now map the different components of these objects (discs/tori, jets, molecular envelope, shocks) with great details down to a few milliarcseconds using either narrowband filters or integral field spectroscopy. One of the nicest illustrations for this is the impressive work by \cite{Gledhill2015}, where they map a spherically symmetric ionised envelope (in Br$\gamma$)  expanding in a bipolar neutral gas envelope (in H$_2$).
I will review here how these advances are helping us to study these objects, which are great test benches for nucleosynthesis and the timing of stellar evolution, the evolution of gas and dust in a harsh radiation field and the understanding of the shaping of non-spherical planetary nebulae.

\section{Finding post-AGB stars}
As the central star quickly shrinks and get hotter during the pos-AGB phase, this phase of stellar evolution is fast, making post-AGB stars rather rare (less than 500 post-AGB stars are known). As the temperature of the central star quickly increase during this phase, post-AGB stars harbour spectral types from B to M. As they are embedded in the dusty envelope formed  during the AGB phase, they are highly obscured. All this makes it very difficult to find post-AGB stars.

The success of the IRAS mission was a great leap forward for the field, as post-AGB stars were identified as luminous object with infrared excess. IRAS colour-colour diagrams (particularly the ([12]-[25] vs [25]-[60]) were thus widely used for follow-up studies to identify stars  located between the AGB and the PN phases. Post-AGB objects were also found by correlating optical catalogues and infrared ones looking for bright optical objects with an infrared excess (\cite[Van de Steene et al. 2000]{Vandesteene2000} and references therein). 

As the post-AGB  objects have circumstellar environment very similar to other dusty objects, confusion can occur, mostly with Young Stellar Objects or massive evolved stars (post-Red SuperGiants). A very useful resource to study post-AGB objects is the online, evolving Torun catalogue of post-AGB objects (\cite[Szczerba et al. 2012]{Szczerba2012}).

\section{Post-RGB stars}
The success of IRAS in the 80s lead to the discovery of many Galactic post-AGBs. The {\it Spitzer Space Telescope} mission  now enables the study of extragalactic post-AGB stars. Studying post-AGB stars in different galaxies can bring a lot of new insights, despite the distances to the objects. With well known  distance to the host galaxies, it is straightforward to determine the luminosities of the studied objects. With nearby galaxies such as the Magellanic Clouds that an overall lower metallicity than the Milky Way, one can also study the effect of metallicity on the late stages of the evolution of low and intermediate mass stars.

A very nice study of post-AGB objects in the Large and the Small Magellanic clouds is presented by \cite{Kamath2014} and \cite{Kamath2015}. Their work combines photometry from Spitzer (to look for IR excess due to dust) and ground-based optical spectroscopy (to suppress contaminants). This led to the identification of 63 post-AGB candidates in the SMC and 154 in the LMC.

As the distances to these objects are well constrained, which is not the case for most of the Galactic post-AGBs, they estimated their luminosities. Surprisingly, most of the objects appear to be emerging from the Red Giant Branch, rather than from the Asymptotic Giant Branch. It is very likely that these object have evolved blue-wards before reaching the AGB phase. \cite{Kamath2016} suggest that these objects are the results of binary interaction, which led to the ejection of the star's envelope before it could reach the AGB phase. They thus  identified a new class of objects, the post-RGB stars. These objects certainly have a Spectral Energy Distribution  (SED) consistent with the presence of a disc due to the presence of a binary. 

The SEDs of these binary objects are different to the ones classically assumed for post-AGB objects with a  double peak (a blue peak due to emission from the star and an infrared peak due to dust emission from the shell). If  a disc is formed via the interaction with the companion, warm dust will  form and be trapped in the disc, leading to a near-infrared excess, filling the gap between the two peak of the SED. Such a shape of the SED is observed for the dusty RV Tau objects (a.k.a. the Van Winckel's stars), which are binary systems on the post-AGB with a stable dusty disc and no visible nebulae (\cite{Vanwinckel2009}).

\section{What can we learn from post-AGB stars studies?}
As we just mentioned, post-AGB objects are not easy to identify but can be key objects to understand stellar evolution and a large variety of physical processes. Determining the chemical composition of their envelopes is easier than on the AGB phase (as molecules and dust make it very difficult to determine the continuum's position in the spectra of  AGBs). As the envelope is ejected after the AGB, post-AGB envelopes' compositions reflect the final chemical enrichment of the gas due to nucleosynthesis. Post-AGB stars are thus ideal test-benches for models of nucleosynthesis. They can also be used to time stellar evolution.

After the AGB phase, the envelope gets detached  and the stars shrinks and becomes hotter. Post-AGB stars are  ideal to study the evolution of gas and dust in a harsh radiation field, and thus better understand the life cycle of matter.

AGB stars are also widely thought to be spherically symmetrical, while many Planetary Nebulae (PNe)  are aspherical. Post-AGB stars can thus be key objects for the study of the onset of asymmetry and better understand the impact of binaries and magnetic field in the shaping of those nebulae.

\subsection{Test for nucleosynthesis and stellar evolution}

As stated before, post-AGBs are ideal to study the products of nucleosynthesis on the AGB (as they suffer from less molecular absorption). Studying post-AGB stars in the SMC and LMC can  help understand the effect of metallicity on nucleosynthesis. For example, a discrepancy appears for Galactic objects between models and observations for the abundance of lead (one of the product of s-process nucleosynthesis on the AGB). \cite{Desmedt2014} have shown that this discrepancy is even larger at low metallicity. Neutron capture with neutron densities between s- and r- neutron capture processes may provide an explanation to this discrepancy (\cite[Lugaro 2015]{Lugaro2015}).
A recent study of post AGB stars in the LMC also confirms  a correlation between the third dredge-up efficiency and the neutron exposure (\cite{Vanaarle2013}).

It has also been shown recently that some meteorites have isotopic ratios consistent with a post-AGB origin (\cite[Jadhav 2013]{ Jadhav2013}) with a composition similar to the prediction for the hydrogen injection phase during a Very Late Thermal Pulse (a thermal pulse during the post-AGB phase).

Finally, and this came out to be one of the most important results from this symposium, the study of post-AGB stars have revealed that  the timing of stellar evolution was certainly wrong. A study of PNe in the Galactic Bulge (\cite[Gesicki 2014]{Gesicki2014}) has shown that
Bloecker's evolution models are too slow to explain the presence of PNe around such low mass stars as found in the Bulge (the models takes too much time to reach a temperature high enough for the gas to be ionised).  The predicted final stellar masses are also  higher than what is measured for White Dwarfs or measurements from asteroseismology. Bloecker's post-AGB evolution models need to be accelerated by a factor of 3. This helps produce new stellar evolutionnary models, that should replace Bloecker's models (\cite{MillerBertolami2016}).

\subsection{Dust evolution after the AGB}
At the end of the AGB, the envelope gets detached, so that the dust goes further away from the star and gets cooler. At the same time, the star becomes hotter and the dust is exposed to a harsher radiation field. The presence of a companion can also lead to the formation of a disc or a torus and dust processing. All these factors lead to a very complex chemistry on the post-AGB, with very complex components observed such as PAHs, fullerenes, and the unidentified 21 microns feature. The most recent knowledge about  dust evolution during the post-AGB phase has been obtained thanks to the study of objects in the Magellanic Clouds. A very thorough work by \cite{Sloan2014} using infrared spectra from the {\it Spitzer Space Telescope} has identified 5 classes of objects clearly clustered  in color-color diagrams. The difference of chemistry is thus likely linked to a difference in density that could be linked to the presence of discs/torii. A morphological study of these objects would be necessary to understand their dust properties.

These studies however enabled a better understanding of the dust/gas composition after the AGB.
For example, the so-called 21 micron feature, an unidentified broad feature seen only in carbon-rich post-AGB stars seems to be very common at low metallicity (\cite[Volk et al. 2011]{Volk2011}) and to be always associated with objects harbouring PAHs in their spectra. Thus, \cite{Cerrigone2011} proposed that it was associated with hydrocarbons. \cite{Sloan2014} showed that it was always associated with two unidentified features at 15.8 and 17.1 microns. They showed that these two features were clearly linked to alkynes, confirming that the 21 micron feature is  associated with hydrocarbons.  Finally, stars harbouring the  21 micron feature have a special class of newly defined PAHs (Sloan et al., 2014), and seem to have a line of sight to the central star is free of dust absorption.
{\it Spitzer} spectroscopy of  post-AGB stars in the Magellanic Clouds also revealed a  new class of PAHs
(\cite[Matsuura et al. 2014]{Matsuura2014}), with a  peak at 7.7 microns. They also showed that the observed class of PAHs for a given object was  linked to its  evolutionary state rather than its metallicity, i.e. that the different kinds of PAHs observed is due to the dust density and central star's radiation field rather than to its initial composition. 

Finally, an unexpected discovery from the {\it Spitzer} surveys of post-AGB stars in the Magellanic Clouds was the presence of very strong and broad emission features around 11.3 microns. A very common feature due to SiC is usually observed at this wavelength but AGB studies have shown that the strength of this feature tend to decrease with metallicity (\cite[Sloan et al. 2006]{Sloan2006}; \cite{Lagadec2007}), as less Si is available in low Z environments to produce silicon carbide. One would thus expect post-AGBs and PNe in the Magellanic Clouds to have a rather weak SiC emission. But these lower metallicity objects harbour more prominent 11.3 microns features than in the Galaxy (where less than a handful of PNe with SiC are known). This 11.3 microns feature could be due to PAHs emission, but its shape is fully consistent with SiC (\cite{Sloan2014}).  SiC can indeed be less abundant but  a bright SiC feature can be observed if  the  condensation sequence is different at low Z, leading to grain coating, with SiC forming on top of amorphous carbon grains (\cite{Sloan2014}; \cite[Lagadec et al. 2007]{Lagadec2007}; \cite[Leisenring et al. 2008]{Leisenring2008}).

\section{Binary post-AGB stars and shaping}
\subsection{Determining the morphology of post-AGB objects}

While most of the stars on the RGB and the AGB are thought to be  more ore less spherical, Planetary Nebulae (PNe) can harbour a wide variety of shapes and be elliptical, bipolar or multipolar. It is also widely accepted that the shaping of these objects is likely to occur during the post-AGB phase and that binarity and magnetic field (sustained by a companion) are important shaping agents.
 
The observed morphology of the  objects appears as projected on the sky, making it difficult to know   their 
intrinsic shape. In  a recent work,  \cite{Koning2013}  constructed a model for post-AGB nebulae  based on a pair of hollow cavities in a spherical dust envelope. They were able to reproduce most of the observed shapes by only changing the orientation and dust densities. One thus has to keep that in mind when trying to study the morphology of an object.

Another difficulty arises when one wants to determine the morphology of an object is the wavelength-dependence induced by radiative transfer effects. This is very well illustrated by a study  of the highly aspherical  post-AGB star HD 161796 by \cite{Min2013}. Using imaging polarimetry, they mapped the dust shell around the object. Polarimetry is a great tool to separate the light scattered by dust  in the envelope and the unpolarised light from the central star. HD 161796 clearly hosts a central density enhancement, which is optically thick at short wavelengths and  thin towards longer wavelengths.
Thus, optical images reveal bright lobes due to scattered light from the torus, while in the infrared one sees emission from this equatorial ring. One thus also has to keep  in mind while studying morphologies that this is something wavelength-dependent, and that to see direct emission from the dust shell one has to observe at mid- or far- infrared wavelengths (see e.g. \cite{Lagadec2011}).

\subsection{Binaries as shaping agents}

  Hydrodynamical models explain many of the
observed aspherical  structures from a structure-magnification mechanism, where a
fast wind from the smaller, hotter, central star  
ploughs into the earlier slow, dense, AGB wind (\cite[Kwok et al. 1978]{Kwok1978})  and amplifies any density asymmetry already present
(\cite[Balick et al. 1987]{Balick1987}): the Generalised 
Interacting Stellar Wind model or GISW. 
Fast, collimated and precessing jets could also explain the formation of multipolar nebulae , not explained by the GISW model
(\cite[Sahai \& Trauger 1998]{Sahai1998}).
  The presence of jets was confirmed by a study of
PPNe by Bujarrabal et al. (2001).They found that in about 80\% of the PPNe, the momentum
of the outflowing material is too high to be powered by radiation pressure only. An extra source of angular momentum is thus needed to explain the presence of these jets.
To explain the formation of jets, two kinds of models have been proposed, implying either a magnetic field or a bianry companion, leading to a long debate in the PN community. Noam Soker and Jason Nordhaus settled the debate by showing that magnetic fields could play an important role, but a single star could not supply enough angular momentum to shape the nebulae: a companion is needed (\cite[Soker 2006]{Soker2006}; \cite[Nordhaus \& Blackman 2006]{Nordhaus2006}).

More and more binary systems are being discovered in
PNe (see the contribution by David Jones to these proceedings (\cite[Jones 2016]{Jones2016}). The detection of binaries in post-AGB systems is made  complex
by  the pulsation of the central stars and their dusty envelopes. Binaries have however been discovered in two emblematic bipolar post-AGBs, OH\,231.8+4.2 (\cite[G\'omez \& Rodr\'iguez 2001]{Gomez2001})  and
the Red Rectangle (\cite[Waelkens et al. 1996]{Waelkens1996}). 
A long lasting quest for binaries in post-AGBs has
been performed by Bruce Hrivnak and his undergraduate students at
Valparaiso University, with radial velocity
monitoring since 1994. They might have detected a binary system with P$>$22 years
(\cite[Hrivnak et al. 2011]{Hrivnak2011}). Binaries are also being discovered by Hans van Winckel and his collaborators, mostly in dusty RV Tau stars, that certainly form a special class of post-AGB stars, as we will discuss now.

\section{Two classes of post-AGB objects}

 It is indeed  important to make the distinction between two kind of observed post-AGB objects. Proto-Planetary Nebulae are post-AGB objects that (as their name indicates) will from PNe, but not all post-AGB objects will form PNe.

\subsection{Proto-Planetary Nebulae}
PPNe have nebulae visible in reflection in the optical and emission in the infrared. The most bipolar PPNe are often associated with 
massive central torii (masses of the order of about a  solar mass or more: \cite[Lagadec et al. 2006]{Lagadec2006}). They have a low
expansion velocity (typically a few km/s: \cite[Peretto et al. 2007]{Peretto2007}. They have a small angular momentum and expand radially (i.e. they have almost no rotation). If the mass loss that created them stops, they will vanish rapidly.

In the recent years, very interesting studies of such objects have been obtained, in particular thanks to new instruments like ALMA, that enable the  study of gas kinematics at very high angular resolution, and thus a better understanding of the formation of torii/jets.
\cite{Sahai2013} resolved a hourglass-shaped torus in the core of the Boomerang nebula using ALMA CO observations. They found this torus to be a hollow cavity with very thick walls of gas and dust. This torus is surrounded by a roughly round, but patchy, high velocity outflow. This outflow appears to be ultra-cool, cooler than the cosmic microwave background, so that they dubbed this object the coolest object in the universe (PPNe are very cool objects to study!).
Using the SMA, \cite{Lee2013} spatially resolved the CO gas around the PPN CRL\,618.
The circumstellar environments consist of a dense equatorial  torus and 
different fast molecular outflows roughly perpendicular to it.
Two episodes of bullet-like ejection are needed to explain these outflows, that clearly are not long lived jets.
Similar structures were found in the Water Fountain Nebula (\cite[Sahai et al. 2016]{Sahai2016}).
The magneto-rotational explosion model of \cite{Matt2006} seems to be a good explanation for these bullet like outflows.

That leads us to the study of magnetic fields at the surface of these objects.
For a thorough review, I will refer the reader to the contribution to this proceedings  by Laurence Sabin.
More and more magnetic fields measurements have been lately obtained  for post-AGB stars.
One of the most impressive one is the detection of a magnetically collimated jet in the water fountain post-AGB  IRAS 15445$-$5449 by \cite{Perez-sanchez2013}. The direction and extent of the jet are clearly coincident with the extended envelope observed in the mid-infrared (\cite[Lagadec et al. 2011]{Lagadec2011}). The short-lived jet  is collimated by a magnetic field of  the order of mG, at $\sim$ 7000 AU from the central star.
The surface magnetic fields of a couple of RV Tau post-AGB stars was also measured by \cite{Sabin2014}.
Their study
suggest that the magnetic field  varies with the pulsation phase of the star and monitoring programs are needed to study the impact of this weak magnetic field on the evolution of the stars.
Finally, as we mentioned before, theoretical models predict that a companion is necessary to sustain a magnetic field strong enough to shape the nebulae. Studying a post-AGB system harbouring a known binary star is thus important to better understand the impact of binaries and magnetic fields on the shaping of nebulae. \cite{Leal-Ferreira2012} confirmed the presence of a magnetic field  around OH231.8+4.2, and measured its strength within a few tens of AU of the binary system, but were not able to determine the morphology of the field. This was done by \cite{Sabin2015} with CARMA observation that showed  polarisation maps indicating an overall organised magnetic field within the nebula, with a toroidal component  aligned with the central equatorial over-density.
This seems to indicate the presence of a  magnetic launching mechanism in the Rotten Egg Nebula.

Recent works are thus showing that the shaping of PPNe is very likely due to the presence of  a companion and  a magnetic field, sustained by this companion.

\subsection{Naked post-AGB stars}
Many  post-AGB stars are also of the dusty RV Tau type, also called naked post-AGB stars, discs post-AGB stars or the Van Winckel's objects. 
They harbour  a near-infrared excess in their SED, due to the presence of dust near the sublimation temperature, in a 
 compact (R$\sim$ 10 AU) disc. This was   confirmed  by their infrared
interferometric measurements (\cite[Deroo et al. 2007]{Deroo2007}). These discs are long lived, as indicated by  crystalline dust, have a very small aperture angle  and a Keplerian  kinematics (see e.g. \cite{Bujarrabal2013}[Bujarrabal et al. 2013]; \cite[Bujarrabal et al. 2015]{Bujarrabal2015}. Their lifetimes are much larger than those of the torii in the core of PPNe.
The photospheres of the central stars are also depleted in refractory elements, as a result of gas and dust separation in the disc, followed by
re-accretion of  the gas. This gas is poor in refractory elements, as these elements with a high condensation temperature are trapped in dust. 
Radial velocities monitoring  indicates that close binary systems (0.5--3 AU) are present in the cores of
those objects (\cite[van Winckel et al. 2009]{Vanwinckel2009}). 
A recent study of one of these objects, IRAS08544-4431, by \cite{Hillen2016}  led to the first image of a post-AGB binary system with  a circumbinary disc.
Using  PIONIER  at the Very Large Telescope Interferometer (VLTI), they obtained an image with a resolution down to the milliarcsec scale and managed to map a circum-companion accretion disc .
This opens a new window to study  binary evolution  and circumstellar disc evolution  in detail over space and over time.

\subsection{Naked vs clothed post-AGBs}
As described in the previous sections, two very distinct classes of post-AGB stars are observed, the dusty RV Taus (Naked post-AGBs)  and the PPNe.
Naked post-AGBs  all  have dusty Keplerian discs (with almost no gas), crystalline dust,  binaries but no long-lived nebulae. They are thus very unlikely to form PNe, as when the central stars will become hot enough for the ionisation to occur, no gas will be present around them. 
PPNe emerging from binary systems have slowly, radially, expanding torii (no Keplerian rotation), and contain amorphous dust and gas. They are short-lived.
One of the explanation for the naked post-AGB stars  is that the configuration of the binary system led to gas accretion from the central star. It  then does not lose mass on the post-AGB and remains cold longer. There are  no ionising photons while gas is present in the circumstellar environment and thus no PN (O. De Marco, private communication).

\section{Conclusion and perpectives}
Post-AGB stars are great tools to test nucleosynthesis and evolution models for stars of low and intermediate masses. The post-AGB phase starting by definition when the envelope of an AGB star gets detached and the star becomes hotter. They are also great test-benches for the study of the evolution of dust in harsh environments of different densities and harbour a very larger variety of complex molecules and dust features such as PAHs, crystaline and amorphous dust.
They are also not easy to identify as the post-AGB phase is short and  the stars are embedded in dust. If they are in close binary systems, the envelopes can be ejected before the AGB phase, leading to the formation of objects of a newly discovered class: the post-RGB stars.
If binary systems manage to evolve till the end of the AGB without ejecting the envelope, they can lead to the formation of bipolar post-AGB stars. Two kinds of aspherical post-AGB object can be formed. PPNe harbour a dense torus and will become bipolar planetary nebulae.
Naked post-AGBs harbour Keplerian discs and no visible nebulae and will not form PNe. 
Finally, the recent advances in observing techniques with high angular resolution now enable us to map circumbinary discs around post-AGB stars and study in real-time the evolution of the discs and the binaries. These high angular resolution techniques also show that, contrarily to what has been thought for decades, the shaping of aspherical nebulae does not necessarily start during the post-AGB phase, as more and more AGB stars appear to be aspherical. This was demonstrated very strikingly by the observations of the nearby AGB star L$_2$ Pup using the extreme adaptive optics system SPHERE on the VLT, which revealed the clear presence of a disc, outflows perpendicular to it and a companion to the central star (\cite{Kervella2015}). We might thus have to study stars on the AGB phase rather than on the post-AGB to understand the beautiful shapes of PNe.

\end{document}